\documentclass[aps,pra,twocolumn,floats,showpacs]{revtex4}
\usepackage{psfig}
\usepackage{amsfonts}

\begin{document}

\title{Entanglement concentration of continuous variable quantum states}
\author{Jarom\'{\i}r Fiur\'{a}\v{s}ek, Ladislav Mi\v{s}ta, Jr.,
and Radim Filip}
\affiliation{Department of Optics, Palack\'{y} University, 17. listopadu 50,
77200 Olomouc, Czech Republic}

\begin{abstract}
We propose two probabilistic entanglement concentration schemes for
a single copy of two-mode squeezed vacuum state. The first scheme is
based on the off-resonant interaction of a Rydberg atom with the cavity
field while the second setup involves the cross Kerr interaction,
auxiliary mode prepared in a strong coherent state and a homodyne
detection. We show that the continuous-variable entanglement
concentration allows us to improve the fidelity
of teleportation of coherent states.
\end{abstract}


\maketitle

\section{Introduction}

Quantum entanglement is an essential ingredient of many protocols for
quantum information processing such as quantum teleportation
\cite{Bennett93,Teleportation} or quantum cryptography \cite{Ekert91}.
In order to achieve optimum performance of these protocols,
the two involved parties, traditionally called
Alice and Bob, should share a pure maximally entangled state.
In practice, however, we are often able to generate only non-maximally
entangled state and, in addition, the distribution of the entangled state
between the two distant parties via some noisy quantum channel will
degrade the entanglement and Alice and Bob will share some partially
entangled mixed state. One of the most important discoveries in the quantum
information theory was the development of the entanglement
purification protocols  that allow Alice and Bob to extract a small number
of highly entangled almost pure states from a large number of weakly
entangled mixed states \cite{Bennett96b,Bennett96,Deutsch96}.
These protocols involve only local operations
and classical communication (LOCC) between the two parties, therefore
they can be performed after the distribution of the entangled states.

In the simplest scenario Alice and Bob share a pure non-maximally
entangled state  in a $d$-dimensional Hilbert space
whose Schmidt decomposition reads
\begin{equation}
|\psi\rangle= \sum_{j=1}^d c_j|\alpha_j\rangle_A|\beta_j\rangle_B,
\label{PSI}
\end{equation}
where each set of states $|\alpha_j\rangle$ and $|\beta_j\rangle$ forms
a basis.
Alice and Bob  would like to prepare from $|\psi\rangle$ a state with
higher entanglement by means of LOCC operations.
Remarkably, this is possible, albeit only with
certain probability, even if they share only a single copy of this
state. The procedure that accomplishes this task was fittingly called
the Procrustean method \cite{Bennett96b}, because it cuts off the Schmidt
coefficients $c_j$ to the size of the smallest one. In this way, Alice and
Bob obtain, with certain probability, a maximally entangled state in
a $d$-dimensional Hilbert space.

In view of the recent interest in quantum information processing
with continuous variables
\cite{Teleportation,Cloning,Cryptography,ErrorCorrection,EntanglementSwapping},
it is highly desirable to establish
experimentally feasible entanglement distillation and concentration
protocols for the continuous variables. Of particular importance are the
protocols for Gaussian states, because  these states can be prepared in
the lab with the use of commonly available resources comprising
lasers, passive linear optics and squeezers (parametric amplifiers).
However, it was proved recently
that it is impossible to distill Gaussian
entangled states by means of Gaussian operations only \cite{Nogo}.
This means that
additional resources beyond the linear optics and homodyne detectors are
required. The distillation protocols for Gaussian states
proposed so far employ the photon-number measurements.
The scheme suggested by Duan {\em et al.} \cite{Duan00} relies on nondemolition
measurement of the total photon number in two (or more) modes and
represents a direct extension of the Schmidt projection method
to infinite-dimensional Hilbert space. The Procrustean schemes considered
by Opatrn\'{y} {\em et al.} \cite{Opatrny00} and further analyzed by
Cochrane {\em et al.} \cite{Cochrane01} are based on
a controlled addition and subtraction of photons. We also note that
several distillation schemes for entangled coherent states
have been proposed \cite{Parker00,Clausen02}.

In this paper, we design two  entanglement-concentration setups for
a single copy of pure two-mode squeezed vacuum state
\begin{equation}
|\psi\rangle= \sum_{n=0}^\infty c_n|n,n\rangle, \qquad
c_n=\sqrt{1-\lambda^2} \lambda^n,
\label{NOPA}
\end{equation}
where $\lambda=\tanh r$ and $r$ is the squeezing constant.
This state can be generated in the process of a non-degenerate
spontaneous parametric downconversion
and provides a common source of the continuous-variable entanglement
in the experiments. The Procrustean procedures that we are
proposing preserve the structure of the state (\ref{NOPA})
while the Schmidt coefficients $c_n$ are transformed to new ones,
$c_n\rightarrow d_n$.
The first scheme is based on a dispersive interaction of a two-level
atom with the microwave cavity field and the atomic-state detection.
The second scheme utilizes a cross Kerr interaction,
coherent states, homodyne
measurements, and linear optics.  The underlying mechanism of both these
schemes is that a certain  auxiliary system experiences a phase shift
that depends on the number of photons in the Alice's mode of the shared
state (\ref{NOPA}). We convert this phase modulation into amplitude
modulation via interference, which allows us to control the amplitude
of the Schmidt coefficients $c_n$. An essential part of our probabilistic
protocols is the measurement on the auxiliary system which tells
us whether the concentration succeeded or failed.

The paper is organized as follows. The first scheme is analyzed in Sec.
II and the second scheme is discussed in Sec. III. Finally, Sec. IV
contains the conclusions.

\section{Entanglement concentration in cavity QED}

\begin{figure}
\centerline{\psfig{figure=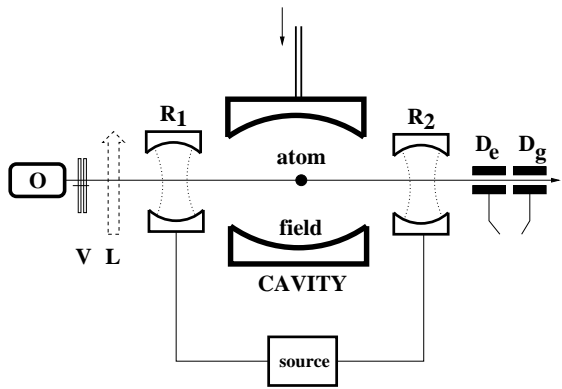,width=0.85\linewidth}}
\caption{Schematic of entanglement concentration setup in cavity QED:
O is atomic oven, V is atomic velocity selector, L is laser excitation
mechanism, R$_1$ and R$_2$ are the Ramsey zones driven by the microwave
source, CAVITY contains the Alice's (Bob's) part
of the entangled state and D$_{e}$, D$_{g}$ are the field ionization
detectors measuring the state of the Rydberg atom.}
\end{figure}

Our first entanglement concentration scheme is designed for the quantum
state of electromagnetic field confined in a high-$Q$ cavity
and is schematically sketched in Fig. 1.
Note, that this setup has been successfully realized experimentally
and employed for the QND measurements of the cavity-field photon number
and the preparation of Schr\"{o}dinger cat states \cite{qnd,Brune96}.
The scheme shown in Fig. 1 is based on an off-resonant
interaction of an (effectively) two-level Rydberg atom with
a single mode of a cavity sandwiched in the Ramsey interferometer.
The atoms are emitted from an oven, their velocity is selected by the
velocity selector, and are excited by a laser pulse to the upper level.
Subsequently, each atom enters the first microwave Ramsey zone
where it is prepared in a coherent superposition of the two (relevant)
long-living circular Rydberg states $| g \rangle$ and $| e \rangle$,
\begin{equation}
|\phi\rangle=\frac{1}{\sqrt{2}} (|  g \rangle + e^{i\varphi_0}
\label{PHI}
|  e \rangle).
\end{equation}
The atom then traverses the cavity that contains the Alice's part
of the shared two-mode state (\ref{NOPA}). The dispersive
atom-field interaction in the cavity is governed by the following
effective Hamiltonian
\begin{equation}
H = \hbar \kappa a^\dagger a \otimes | e \rangle\langle  e |,
\label{H}
\end{equation}
where $a$ is annihilation operator of Alice's mode and
$\kappa$ is effective atom-field interaction constant.
The coupling (\ref{H}) results in a phase shift
$\Delta\varphi=a^{\dagger}a\varphi$ of the state $| e \rangle$ that is
linearly proportional to the number of photons
in the mode $A$. On the other hand,  the state $| g \rangle$ is not changed
by the interaction. The single-photon
phase shift $\varphi=\kappa t$, where $t$ is an
effective interaction time, can be adjusted to the required
value by a proper selection of the atomic velocity.

\begin{figure}
\centerline{\psfig{figure=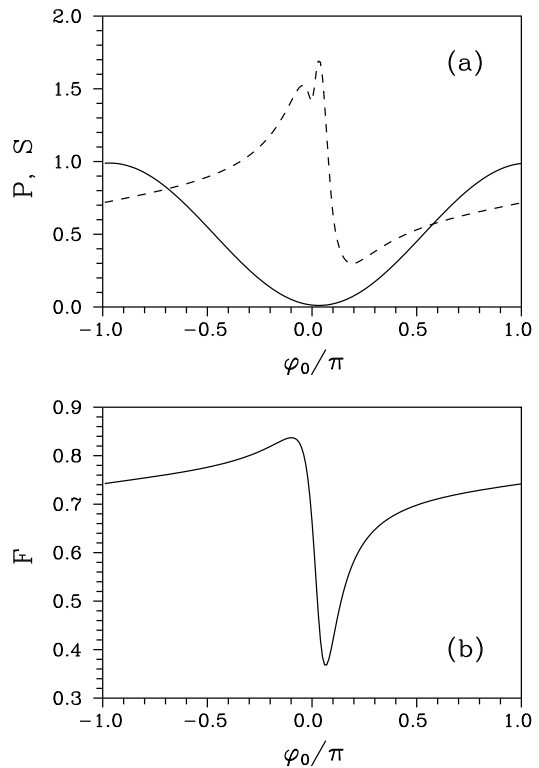,width=0.8\linewidth}}
\caption{The performance of the entanglement concentration scheme shown in
Fig. 1 for $\lambda=1/2$
and $\varphi=\pi/10$. (a) Probability of success $P$ (solid line)
and the von Neumann entropy $S$ after the concentration (dashed line)
and (b) the fidelity $F$ of teleportation of
coherent states are plotted as functions of $\varphi_0$.
For the input state, $S_{\rm in}=0.75$ and $F_{\rm in}=0.75$.}
\end{figure}

After leaving the cavity, the atom passes through the
second microwave Ramsey zone, where it undergoes a $\pi/2$ Rabi rotation,
\begin{equation}
|  g \rangle \rightarrow
\frac{1}{\sqrt{2}}(|  g \rangle+|  e \rangle),
\qquad
|  e \rangle \rightarrow
\frac{1}{\sqrt{2}}(|  e \rangle-|  g \rangle).
\label{HADAMARD}
\end{equation}
The resulting state of the atom and the two-mode field reads
\begin{eqnarray}
|\Psi\rangle =\frac{1}{2}\sum_{n=0}^\infty c_n
(1-e^{i\varphi_0-i n \varphi})|  g \rangle\otimes |n,n\rangle
\nonumber \\
+\frac{1}{2}\sum_{n=0}^\infty c_n
(1+e^{i\varphi_0-in \varphi})|  e \rangle \otimes |n,n\rangle.
\end{eqnarray}
To complete the procedure, we measure the state of the atom by means of
state-selective ionization detectors exhibiting almost unit detection
efficiency. The entanglement concentration succeeds only if the
atom is found to be in the ground state $|  g \rangle$.
The new Schmidt coefficients after this conditional transformation
read
\begin{equation}
d_{n}= i c_n \exp\left(i\frac{\varphi_0-n\varphi}{2}\right)
\sin\left(\frac{n\varphi-\varphi_0}{2}\right).
\end{equation}

The irrelevant overall phase factor $i\exp(i\varphi_0/2)$ can be dropped.
Moreover, the phase factor $\exp(-i n\varphi/2)$
can easily be compensated by appropriate
phase shift or simply by properly redefining the quadratures of the
Alice's mode. After these transformations, the
new Schmidt coefficients become real and after renormalization we get
\begin{equation}
d_n= \sqrt{\frac{1-\lambda^2}{P}} \, \lambda^n \,
\sin\left(\frac{n\varphi-\varphi_0}{2}\right),
\label{DN}
\end{equation}
where
\begin{equation}
P=\frac{1}{2}-\frac{1-\lambda^2}{2}
\frac{\cos(\varphi_0)-\lambda^2\cos(\varphi+\varphi_0)}{1-2\lambda^2\cos(\varphi)+\lambda^4}
\end{equation}
is the probability of success of the conditional transformation.
Clearly, two trends are competing in Eq. (\ref{DN}).
The exponential decay $\lambda^n$ is for certain $n$ partially
compensated by the second term $\sin[(n\varphi-\varphi_0)/2]$ which grows
with $n$ up to $n_{\rm max}=(\pi+\varphi_0)/\varphi$.
This allows us to increase the entanglement of the shared state.

Formally, the conditional transformation can be described as
a diagonal filter applied to the input two-mode density matrix $\rho_{AB}$.
Define operator
\begin{equation}
A=\sum_{n=0}^\infty \sin\left(\frac{n\varphi-\varphi_0}{2}\right)
|n\rangle\langle n|.
\end{equation}
The output (unnormalized) density matrix is given by
\begin{equation}
\rho_{\rm out}= A\otimes \openone_B \rho_{AB}A^\dagger \otimes\openone_B,
\end{equation}
where $\openone_B$ stands for an identity operator on the Hilbert space
of the Bob's mode.

Since the conditional transformation preserves the purity of the
two-mode state, we can conveniently quantify the entanglement as
the von Neumann entropy of the reduced density matrix of the Alice's mode,
\begin{equation}
S=- \sum_{n=0}^\infty |d_n^2| \ln |d_n^2| .
\label{S}
\end{equation}
The entropy $S$ is plotted in Fig. 2(a) as a function of $\varphi_0$ for
fixed $\varphi$ and $\lambda$. Before concentration, Alice and Bob share
two-mode squeezed vacuum (\ref{NOPA}), and the entropy (\ref{S})
reads
\begin{equation}
S=-\ln(1-\lambda^2)-\frac{\lambda^2}{1-\lambda^2}\ln \lambda^2.
\end{equation}
For the data in Fig. 2, we obtain $S_{\rm in}=0.75$.
The figure 2(a) clearly shows that for certain interval of phase shifts
$\varphi_0$ our procedure allows us to conditionally increase the amount of
entanglement in the pure two-mode state shared by Alice and Bob.

Let us now
demonstrate that the entanglement concentrated in this way is useful in
practical tasks. To be specific, we consider the teleportation of
coherent states in the Braunstein-Kimble scheme \cite{Teleportation}
where our state is used as the quantum channel.
Making use of the transfer operator
formalism \cite{Hofmann00,Kurzeja02},
we can express the fidelity of teleportation as follows,
\begin{equation}
F= \frac{1}{2} \,\sum_{m=0}^\infty \,\sum_{n=0}^\infty
 {m+n \choose n} \frac{d_m d_n^\ast}{2^{m+n}} .
 \label{FTELE}
\end{equation}
On inserting the Schmidt coefficients (\ref{DN}) into Eq. (\ref{FTELE})
and carrying out the summations we obtain analytical formula for
the fidelity of teleportation of coherent states,
\begin{equation}
F= \frac{1-\lambda^2}{4P} \left[
\frac{1}{1-\lambda\cos(\varphi/2)} -
\frac{\cos(\varphi_0)-\lambda\cos(\varphi/2+\varphi_0)}%
{1-2\lambda\cos(\varphi/2)+\lambda^2}
\right].
\label{FTELEANALYTIC}
\end{equation}
The fidelity $F$ is plotted in Fig. 2(b). For fixed $\lambda$ and $\varphi$
we can optimize the phase $\varphi_0$ so that the teleportation fidelity $F$
will be maximized. For the data used in Fig. 2, we find that it is
optimum to set $\varphi_0\approx-\pi/10$, which yields the fidelity
$F=0.837$ and the probability of success is $P=0.05$.
This should be compared with the fidelity $F_{\rm in}=0.75$
that is achieved when the original two-mode squeezed
vacuum with $\lambda=1/2$  serves as the quantum channel.
This improvement in fidelity is quite significant and clearly illustrates
the practical utility of our procedure.

\begin{figure}
\centerline{\psfig{figure=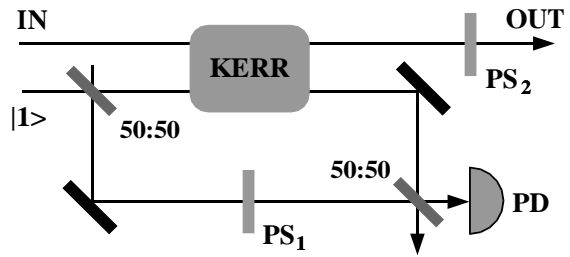,width=0.85\linewidth}}
\caption{Optical implementation of the entanglement concentration scheme
shown in Fig. 1.}
\end{figure}

\section{Entanglement concentration for traveling light fields}

To implement the scheme discussed in the preceding Section
for traveling light fields, we could replace the atomic Ramsey interferometer
with a Mach--Zhender interferometer for a single photon and couple this
auxiliary photon to the Alice's mode via nonlinear medium with cross
Kerr effect, see Fig. 3. A similar setup has been proposed by Gerry for
the generation of Schr\"{o}dinger cat states \cite{Gerry00}.
However, this scheme has several drawbacks. First, the currently
achievable Kerr nonlinearities are rather low. Secondly, we have to
prepare a single photon.  Therefore we propose an alternative
scheme,  see Fig. 4. In that setup, an auxiliary mode $C$ is
prepared in a (strong) coherent state $|\alpha\rangle$ and then
interacts with the Alice's mode $A$ in the Kerr medium described by the
Hamiltonian
\begin{equation}
H_{\rm Kerr}= \hbar \kappa a^\dagger a c^\dagger c.
\end{equation}
After the interaction, we project the output state into coherent state
$|\beta\rangle$ in the eight-port homodyne detector.

\begin{figure}[!t!]
\centerline{\psfig{figure=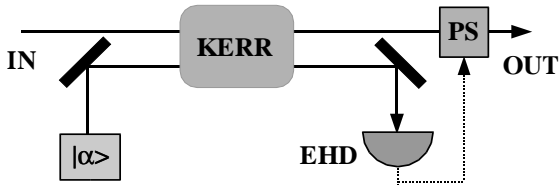,width=0.85\linewidth}}
\caption{Schematic of the entanglement concentration setup
for traveling light fields that is based on auxiliary coherent states,
cross Kerr interaction, eight-port homodyne detection (EHD),
and a linear phase shift depending on the outcome of
the measurement (PS).}
\end{figure}

The principle of the operation of this scheme may be explained as
follows. If there are $n$ photons in the mode $A$, then the coherent
state $|\alpha\rangle$ evolves to $|\alpha e^{in\varphi}\rangle$, where
$\varphi=-\kappa t$ and $t$ is the effective interaction time. The
probability of projecting into $|\beta\rangle$ is given by
\begin{equation}
P(\beta|n)=\frac{1}{\pi} |\langle \beta|\alpha e^{in\varphi}\rangle|^2.
\end{equation}
From this formula we can see that the probability may grow with $n$
if $\beta$  belongs to certain region of the phase space.
Without loss of generality, we may assume that $\alpha$ is real and
positive and define $\beta=|\beta|\exp(i\varphi_0)$.
After projecting into $\beta$, the new Schmidt coefficients can
be expressed as
\begin{equation}
d_n \propto c_n \langle \beta| \alpha e^{in\varphi}\rangle.
\end{equation}
Making use of the formula for the scalar product of two coherent states
\cite{Perina91}
\begin{equation}
\langle
\beta|\alpha\rangle=
\exp\left(-\frac{1}{2}|\alpha|^2-\frac{1}{2}|\beta|^2 +\beta^\ast\alpha\right)
\end{equation}
we obtain
\begin{equation}
d_n \propto  c_n \exp(q_n+i\phi_n),
\end{equation}
where
\begin{eqnarray}
q_n &=& |\alpha \beta | \cos(n\varphi-\varphi_0),
\label{QN} \\
\phi_n &=& |\alpha\beta|\sin(n\varphi-\varphi_0).
\label{PHIN}
\end{eqnarray}

\begin{figure}[!t!]
\centerline{\psfig{figure=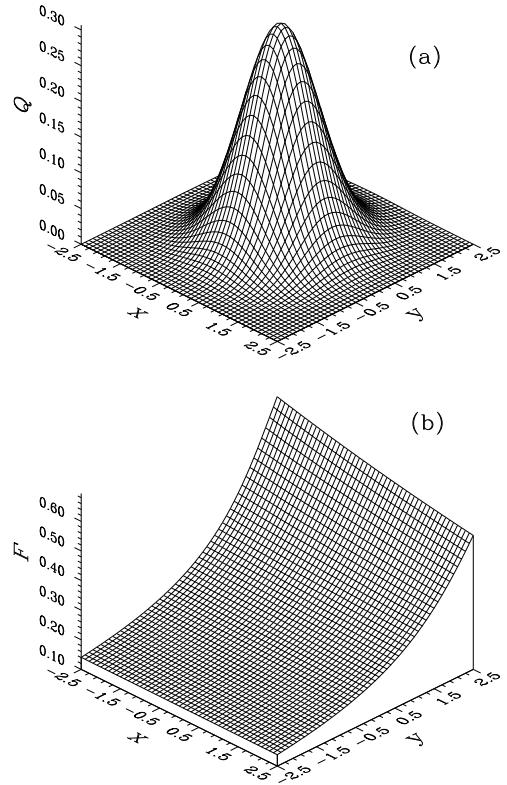,width=0.75\linewidth}}
\caption{(a) The Q-function $Q(\beta)$ of the output state of the
auxiliary mode and (b) the overlap ${\cal{F}}(\beta)$ are shown for
$\lambda=1/2$, $\alpha=10$, $\varphi=\pi/100$, and $N=10$.
The coordinates $x$ and $y$ are defined as $x+iy=\beta-\alpha$.}
\end{figure}

The old coefficients $c_n$ are amplified or de-amplified by the factor
$\exp(q_n)$. The highest enhancement occurs for $n=\varphi_0/\varphi$,
when $q_n=|\alpha\beta|$. For strong auxiliary signal $\alpha \gg 1$, the
phase $\varphi_0$ between $\beta$ and $\alpha$ will typically be of
the order $|\varphi_0| \approx 1/|\alpha|$ and also
$|\beta|\approx|\alpha|$ will hold. Since the nonlinear phase shift
$n\varphi$  will typically be very small for all $n$ for which $c_n$
substantially differs from zero, $n\varphi\ll 1$, we can expand the
expressions (\ref{QN}) and (\ref{PHIN}) in Taylor series and keep
only terms up to linear in  $n\varphi$,
\begin{eqnarray}
q_n&=& |\alpha\beta| \cos(\varphi_0) +n\varphi|\alpha\beta|\sin(\varphi_0),
\nonumber \\
\phi_n&=& -|\alpha\beta|\sin(\varphi_0) +n\varphi|\alpha\beta|\cos(\varphi_0).
\label{TAYLOR}
\end{eqnarray}
This approximation is valid when
\begin{equation}
|\alpha|n\varphi\ll 1
\label{LINEAR}
\end{equation}
holds. Within this approximation the exponents $q_n$ are
linearly proportional to $n$ and  the conditional transformation
preserves the structure of the two-mode squeezed state:
\begin{equation}
|d_n| \propto \tilde{\lambda}^n, \qquad
\tilde{\lambda}=\lambda \exp(\varphi|\alpha\beta|\sin\varphi_0).
\end{equation}
Since $|\beta|\sin\varphi_0 \lesssim 1$, it is the product $\varphi|\alpha|$
that determines the modulation of the input Schmidt coefficients.
A weak Kerr nonlinearity (small  phase shift $\varphi$) can be
compensated by using a sufficiently strong auxiliary coherent
state $|\alpha\rangle$. Furthermore, the strength of the cross-Kerr
interaction can be enhanced by many orders of magnitude in a coherently
prepared resonant atomic medium. A medium with electromagnetically
induced transparency can exhibit an extremely large Kerr nonlinearity
\cite{Schmidt96,Imamoglu97,Hau99,Kash99,Lukin00} that would suffice for
the practical implementation of the present entanglement concentration scheme.

One undesirable effect of the projection into coherent state
$|\beta\rangle$ is the  phase modulation $\phi_n$ of
the Schmidt coefficients. Note, however, that if the approximation
(\ref{LINEAR}) holds then it follows from Eq. (\ref{TAYLOR}) that the
conditional phase shift $\phi_n$ is linearly proportional to $n$
and can thus be removed by a suitable phase shifter PS.
The actual phase shift is proportional to the real part of $\beta$
and we must use a feedforward scheme, where the operation of the
PS (e.g., a Pockels cell) is controlled by the measurement outcome,
as is schematically indicated in Fig. 4.

\begin{figure}[!t!]
\centerline{\psfig{figure=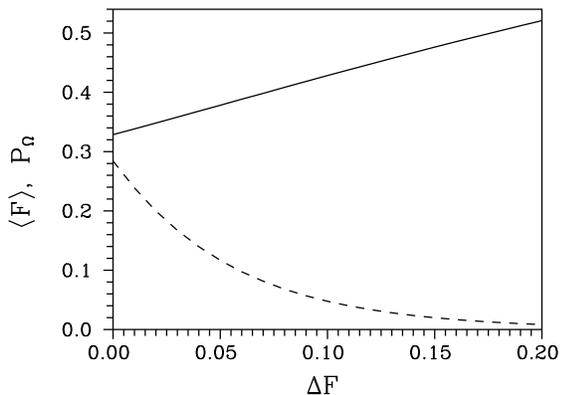,width=0.85\linewidth}}
\caption{The average fidelity $\langle {\cal{F}}\rangle$ (solid line)
and the probability of success (dashed line). All the parameters are
the same as in Fig. 5.}
\end{figure}

After this qualitative discussion, let us now provide a rigorous
mathematical description of the entanglement concentration scheme shown
in Fig. 4. The probability that $\beta$ will be measured in the EHD is
the sum of all the conditional probabilities $P(\beta|n)$ multiplied by
the probabilities $|c_n|^2$ that there are $n$ photons in the mode $A$,
\begin{equation}
Q(\beta)= \frac{1-\lambda^2}{\pi}\sum_{n=0}^\infty \lambda^{2n}
\exp(-|\alpha e^{in\varphi}-\beta|^2).
\label{QBETA}
\end{equation}
The normalized Schmidt coefficients corresponding to the measurement outcome
$\beta$ read
\begin{equation}
d_n(\beta)= \frac{\sqrt{1-\lambda^2} \lambda^n \exp\left[
\alpha\beta^\ast e^{in\varphi}-in\varphi |\alpha\beta|\cos\varphi_0\right]}
{\sqrt{\pi Q(\beta)}\exp\left(|\alpha|^2/2+|\beta|^2/2\right)}.
\label{DBETA}
\end{equation}
We need to establish a criterion according to which we will accept
or reject the state in dependence on the measurement outcome $\beta$.
The most natural approach is to choose some reasonable figure of merit
${\cal{F}}(\beta)$ that has to be evaluated for each $\beta$ and then
specify a region $\Omega$ in the phase space where this figure of
merit is sufficiently large. The entanglement concentration succeeds
only if $\beta\in\Omega$  and fails otherwise.
It follows that the concentration will yield a mixture of the states
\begin{equation}
|\psi(\beta)\rangle=\sum_{d=0}^\infty d_n(\beta)|n,n\rangle
\end{equation}
and the density matrix of the output state shared by Alice and
Bob can be expressed as follows,
\begin{equation}
\rho_\Omega= \frac{1}{P_\Omega}\int_\Omega d^2\beta \, Q(\beta)
|\psi(\beta)\rangle\langle\psi(\beta)|.
\end{equation}
Here
\begin{equation}
P_\Omega= \int_\Omega d^2 \beta \, Q(\beta)
\label{POMEGA}
\end{equation}
denotes the probability of success of the concentration, i.e.,
the probability that the measurement outcome $\beta$ will belong to $\Omega$.

Different figures of merit ${\cal{F}}(\beta)$ may be suitable
depending on the
intended usage of the shared quantum state. For instance, if that state
shall be used to teleport coherent states, then it will be natural to
employ the fidelity (\ref{FTELE}) as the figure of merit.
Here we make use of an even simpler quantity, namely the fidelity
\begin{equation}
{\cal{F}}(\beta)=\left|\langle\Phi_N|\psi(\beta)\rangle\right|^2
\end{equation}
between
the conditionally prepared state and a maximally entangled state in
the Hilbert space of the first $N+1$ Fock states,
\begin{equation}
|\Phi_N\rangle=\frac{1}{\sqrt{N+1}} \sum_{n=0}^N |n,n\rangle,
\label{NMAX}
\end{equation}
The fidelity depends on the sum of the first $N+1$ Schmidt coefficients
and is explicitly given by the following formula,
\begin{equation}
{\cal{F}}(\beta)= \frac{1}{N+1}\left|\sum_{n=0}^N d_n(\beta)\right|^2.
\end{equation}

\begin{figure}[t]
\centerline{\psfig{figure=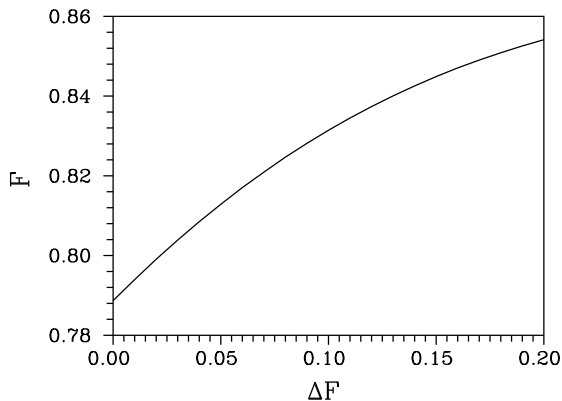,width=0.85\linewidth}}
\caption{The fidelity of the teleportation of coherent states when
the state $\rho_\Omega$ after  concentration is used as the
quantum channel. All parameters are the same as in Fig. 5. }
\end{figure}

In the rest of this section we present the results of the numerical
calculations for $\alpha=10$, $\varphi=\pi/100$, and $\lambda=1/2$.
The $Q$-function (\ref{QBETA}) of the output auxiliary mode is plotted in Fig.
5(a). Since the assumed nonlinear single-photon phase shift $\pi/100$
is relatively small, the $Q$-function is practically identical to the input
Gaussian $Q_{\rm in}(\beta)=  \exp(-|\alpha-\beta|^2)/\pi$.
The function ${\cal{F}}(\beta)$ is plotted in Fig. 5(b) for $N=10$.
We can see that there are regions in the phase space where
$\cal{F}$ is higher than the fidelity corresponding to the input
two-mode squeezed vacuum, ${\cal{F}}_0=0.273$.
As described above, we define  $\Omega$ as the region of the phase space
where ${\cal{F}}(\beta) \geq {\cal{F}}_0+\Delta {\cal{F}}$.
The dependence of the average fidelity
\begin{equation}
\langle {\cal{F}} \rangle = \frac{1}{P_\Omega}
\int_{\Omega} d^2\beta\, Q(\beta) {\cal{F}}(\beta)
\end{equation}
on  $\Delta {\cal{F}}$ is plotted in Fig. 6. Also the probability
of successful entanglement concentration (\ref{POMEGA}) is shown there.
As the gap $\Delta\cal{F}$ becomes larger, the average fidelity
increases while the probability decreases.

Finally, we show that the entanglement concentrated in this way is
suitable for the teleportation. We can calculate the average
teleportation fidelity similarly as $\langle{\cal{F}}\rangle$.
We evaluate the fidelity of teleportation $F(\beta)$ for each
$\beta\in \Omega$ by inserting the relevant Schmidt coefficients
$d(\beta)$ given by Eq. (\ref{DBETA}) into Eq. (\ref{FTELE}) and
then we average $F(\beta)$ over $\Omega$ with the properly normalized
probability density $Q(\beta)/P_\Omega$.
The results are shown in Fig. 7. We can see that the
teleportation fidelity monotonically grows with $\Delta \cal{F}$
and for all $\Delta {\cal{F}}>0$  we have $F>0.75$.  This example
confirms that our procedure indeed extracts more useful entanglement
from the input two-mode squeezed vacuum.

\section{Conclusions}

In this paper, we have designed two schemes for probabilistic
concentration of continuous-variable entanglement. The Procrustean
protocols that we are proposing have the important property that they
can be applied several times to a single copy of the
shared two-mode entangled state. Thus we could, in principle,
extract a state with very high entanglement, at the expense of
a low probability of success. When repeating the concentration
procedure, one could optimize the relevant  parameters such as the
phase shifts $\varphi_0$ and $\varphi$
in order to achieve the optimum performance of the schemes.
In view of the recent advances in cavity-QED experiments and
the preparation of media with extremely high Kerr nonlinearity,
we may hope that the schemes proposed in the present paper will become
experimentally feasible in a near future.

\acknowledgments

This work was supported by the EU grant under QIPC project
IST-1999-13071 (QUICOV) and project LN00A015 of the Czech Ministry of
Education.

\end{document}